\documentclass[preprint2]{aastex}
\input epsf.sty
\usepackage{natbib}

\shorttitle{The Tully-Fisher relation for Hickson Compact Groups}
\shortauthors{Mendes de Oliveira et al.}

\begin{document}

\title{Dynamical effects of interactions and the \\
    Tully-Fisher relation for Hickson compact groups
}

\author{C. Mendes de Oliveira \altaffilmark{1,2,3},
P. Amram \altaffilmark{4}, H. Plana \altaffilmark{5} \and C.
Balkowski \altaffilmark{6}}

\altaffiltext{1}{Universidade de S\~ao Paulo, IAG, 
Departamento de Astronomia, Rua do Mat\~ao 1226, 
05508-900, S\~ao Paulo, SP, Brazil}

\altaffiltext{2}{Max-Planck-Institut f\"ur Extraterrestrische Physik,
Giessenbachstrasse, D-85748 Garching b. M\"unchen, Germany}

\altaffiltext{3}{Universit\"ats-Sternwarte der LMU, 
Scheinerstrasse 1, D-81679 M\"unchen, Germany}

\altaffiltext{4}{ Observatoire Astronomique Marseille-Provence \&
Laboratoire d'Astrophysique de Marseille,  2 Place Le Verrier,
13248 Marseille Cedex 04, France}

\altaffiltext{5}{Universidade Estadual de Santa Cruz, 
Dept. Ciencias Exatas e Tecnologicas, 
Rodovia Ilheus-Itabuna, km. 16, 
45650-000, Ilh\'eus, BA, Brazil}

\altaffiltext{6}{Observatoire de Paris, GEPI, CNRS and Universite
Paris 7, 5 Place Jules Janssen, F-92195 Meudon Cedex, France}

\email{oliveira@astro.iag.usp.br}


\begin{abstract}

We investigate the properties of the $B$-band Tully-Fisher (T-F)
relation for 25 compact group galaxies, using $V_{\rm max}$ derived from
2-D velocity maps.  Our main result is that the majority of the Hickson
Compact Group galaxies lie on the T-F relation. However, about 20\% of the
galaxies, including the lowest-mass systems, have higher $B$ luminosities
for a given mass, or alternatively, a mass which is too low for their
luminosities. We favour a scenario in which outliers have been brightened
due to either enhanced star formation or merging. Alternatively, the
T-F outliers may have undergone truncation of their dark halo due to
interactions. It is possible that in some cases, both effects contribute.
The fact that the $B$-band T-F relation is similar for compact group and
field galaxies tells us that these galaxies show common mass-to-size
relations and that the halos of compact group galaxies have not been
significantly stripped inside $R_{25}$.  

  We find that 75\% of the compact
group galaxies studied (22 out of 29) have highly peculiar velocity
fields. Nevertheless, a careful choice of inclination, position angle
and center, obtained from the velocity field, and an average of the
velocities over a large sector of the galaxy enabled the determination
of fairly well-behaved rotation curves for the galaxies.  However,
two of the compact group galaxies which are the most massive members
in M51--like pairs, HCG 91a and HCG 96a, have very asymmetric rotation
curves, with one arm rising and the other one falling, indicating,
most probably, a recent perturbation by the small close companions.

\end{abstract}




\section{Introduction}

 The influence of environmental effects on the internal dynamics and
matter distribution  of compact group galaxies has not yet been
clearly established, mostly due to lack of reliable kinematic
data. A recent study of the internal kinematics of 30 galaxies by
Nishiura et al. (2000, hereafter N2000) found that asymmetric and
peculiar rotation curves are more frequently seen in the Hickson
Compact Groups (HCG) spiral galaxies than in field or cluster
spirals and the dynamical properties of the galaxies do not seem
to correlate with any group or galaxy parameter.  An older but
very influential study is that by Rubin et al.
(1991, hereafter R1991), who analyzed rotation curves for 32
Hickson compact group galaxies and found that 2/3 of them had
peculiar rotation curves. For the subsample of galaxies for which
rotation curves were possible to be derived, they found a large
offset of the T-F relation with respect to the field
relation in the sense that galaxies in compact groups have ``too
low velocities for their luminosities or, alternatively,
luminosities which are overbright for their rotation velocities.''
This suggested to the authors
 that spiral galaxies in
compact groups have low mass-to-light ratios compared to field galaxies by
about 30\% which, in turn, could be explained if compact group galaxies
have smaller dark halos than their field counterparts.  Given that
compact groups are environments where tidal encounters are common,
it may be expected that interactions have stripped or disrupted
the galaxy dark halos at some level.  These conclusions have important
consequences for the determination of group lifetimes and understanding
how compact groups evolve and eventually merge.

We revisit this important problem using a new dataset for 25
galaxies, of which 13 are in common with either the sample
of R1991 or N2000 or both.  In this
paper we re-examine the T-F relation for compact group
galaxies. Our study differs from the previous work in that it uses rotation
curves obtained from 2-dimensional velocity fields. In a number of cases
this allows a more accurate determination of the rotation curves than was
possible with slit spectroscopy. Additionally, a fuller characterization of
the kinematics of each galaxy is possible.  
We show that, with these new data, the
T-F relation for compact group galaxies is similar to  that
for galaxies in less dense environments.
The exceptions to this conclusion are seen 
for some of the least massive galaxies in our sample.

This paper is organized as follows. In Section 2 we present the set of
rotation curves that are used.  In Section 3 we illustrate and discuss
comparisons of rotation curves for several galaxies in common with R1991
and N2000 and we show that 2-D spectroscopy is
needed if one wants to accurately describe the kinematics of interacting
galaxies.  Sections 4 and 5 show the results on the T-F 
relation and a discussion
respectively.

\section{Data}

  The data used in this paper are gathered from three publications
which studied the kinematics of galaxies in nine  compact groups (HCG 10, 16,
19, 87, 88, 89, 90, 91, 96, 100).
A detailed
description of the observations and data reduction can be found
in Mendes de Oliveira et al. (1998), Plana et al. (2003) and
Amram et al. (2003). In summary, our dataset consists of H$\alpha$
emission-line velocity
maps obtained with a Fabry-Perot system, from which rotation curves were
obtained.
For the T-F study we excluded from the sample
all the elliptical galaxies. We also excluded galaxies
HCG 10a, 10c, 16d and 100a because their rotation curves have
a very short extension, well short of R$_{25}$ and galaxies HCG 16b
and HCG 100b due to their extremely peculiar rotation curves.
We included HCG 19a in our
sample, although Hickson (1993) classifies it as an E2 galaxy, because
we have reclassified it as an S0 based on its kinematic properties.
In addition, we included in our sample unpublished data for
two other galaxies (HCG 07c and  HCG 79d).

  We show in Figs. 1a and 1b the rotation curves for the galaxies
in the sample studied in this paper. The x-axis plots r/R$_{25}$,
the galaxy radius along its major axis normalized by R$_{25}$,
the length of the
major axis at the  25 mag arcsec$^{-2}$ isophote (as given by Hickson
1993; note, however that in Plana et al. 2003 and Amram et al. 2003,
the value for R$_{25}$ was taken from de Vaucouleurs et al. 1991, the 
RC3, resulting in slightly
different numbers from those used here).
In order to obtain a V$_{max}$ for use in the T-F relation (represented by
a black dot in each subpannel of Figs. 1a and 1b) we have computed the maximum
velocity of the average velocity curve (single average of the values for
the receding and approaching sides). V$_{max}$ is not the best
kinematic parameter  to be used in the T-F relation, since
V$_{flat}$, the velocity of the flat portion of the curve or V$_{2d}$,
the velocity at two times the galaxy scale length are known to yield
less scatter in the T-F relation. However, the use of V$_{max}$ allowed us to
use the largest possible number of compact group galaxies. 

  The control sample used for the T-F comparison was the
sample of Sb and Sc galaxies from Courteau (1997), where we used
their ``Vmax'' as the equivalent to the maximum velocity determined by us.
We also used the sample of galaxies in the Ursa Major cluster
(Verheijen 2001), in order
to fill in the lower end of  the mass sequence  in the T-F diagram. We
note that our values of V$_{max}$ are adjusted by the cosmological correction
(1+z) as are also the values given for the galaxies in the control samples.
All velocities were corrected for the same Virgocentric infall
(Paturel et al. 1997) when distances were obtained using
the Hubble law (for our sample and Courteau's sample).

  We note that in Fig. 1 the rotation of HCG 87a is one-sided because
half of this almost edge-on galaxy is not observed due to a strong dust
lane. 

\section{Comparisons with previous long-slit spectroscopy}

  The 2-D velocity maps obtained with a Fabry-Perot instrument
allowed us to mimic a slit through our data in order to recover  the
rotation curves obtained from long-slit spectroscopy.  In this way, we
could make a direct comparison with previous rotation curves obtained by
other authors in order to investigate the reason for existing differences
in the shapes and amplitudes of the rotation curves between different
studies, for a given galaxy.

  We give an example in Fig. 2, where it is clear why with
long-slit spectroscopy the derived rotation curves do  not always
correspond to the overall kinematics of the galaxies.  The rotation curve
derived by R1991 for HCG 88a (reproduced in Fig. 2) shows asymmetries
with disagreement between the two sides of the galaxy (a bifurcation
of the curve at a radius of 10 arcsec).  We overplot on Fig. 2 the
Fabry-Perot data for HCG 88a, with values restricted to a narrow cone
around the galaxy major axis (to mimic a slit), using the particular
values of center, inclination and position angle derived by R1991
photometrically.  Inspecting the figure we see that for radii larger
than 10 arcsec, there is good agreement between the slit-spectroscopy
and Fabry-Perot data, if the photometrically derived center, inclination
and position angle are used.  However, the overall shape of the curve
changes when such parameters are derived from the velocity map and a large
sector of the galaxy is averaged (the resulting rotation curve is shown
in Plana et al. 2003 and in Fig. 1). In particular, if the kinematic
parameters are used and velocities over a large sector of the
galaxies are averaged, the bifurcation present in Fig. 2, at r $\sim$
10 arcsec, disappears, and the galaxy has normal kinematics, with both
sides of the curve matching.  

In order to understand the magnitude of the variations of the parameters
measured from the kinematic and photometric data we list in Table 2 values
for the inclination, PA and differences between centers measured from the
continuum images and velocity fields.  Detailed discussion on the method
used to obtain the center, PA of the major axis, and inclination of the
galaxies are given in Amram et al. (1996).  We included in the table not
only the galaxies used in the Tully-Fisher relation but also those that
were considered too peculiar or for which the gas extension was too short.
We list in the last four columns the values of inclination and PA given
by R1991 and N2000, when available.

  A comparison of the inclinations derived from the Fabry-Perot map
and from the continuum images was done by subtracting columns 2 and 3
of Table 2 (and normalized to one of the two).  The differences spread
approximately around zero, with an rms of 20\% (for the galaxies used in
the Tully-Fisher relation). This indicates that there is no systematic error
introduced in the determination of the maximum velocities due to the
measurement of the inclination from either photometric or kinematic data.
However, a similar comparison with the smaller subsample in common with R1991
and N2000 show a slight overestimate of the inclinations obtained by
these authors, as compared to our estimates. This would indicate that
the V$_{max}$ derived by them would be expected to be lower on average
than those obtained by us (and in fact, the values of V$_{max}$ derived by
R1991 in particular tend to be lower than those derived in our study).

   A comparison between columns 4 and 5 of Table 2, of the kinematic and
photometric PA's for the galaxies used in the T-F analysis, shows that
about 1/3 of the galaxies have misalignments (greater than 10 degrees)
between the kinematic and photometric axes. This may be an indication that
the gas is not in equilibrium in these galaxies (HCG 10d, 16c, 19a, 88b,
89d, 91a, 91c, 96a).  If that is the case, and for an example the gas
is collapsing and/or there is additional dispersion, then the V$_{max}$
we compute using the rotation curve  of the gas is a lower limit to
the real V$_{max}$ (indicated by the total mass).  In other words,
the real mass of the galaxy would then be higher than what we infer
from our measurements. We note that four of the six galaxies that were
too peculiar or had too short gas extensions to enter the T-F analysis
also present misalignments between the kinematic and photometric axes
(HCG 16b, 16d, 10c, 100b).

   Similar results are obtained
when comparing the kinematic PA's to the corresponding photometric
values derived from either R1991 and/or N2000 for the smaller subsample
of galaxies in common (comparing columns 4 and 8 and columns 4 and 10
of Table 2).

  We are not able to make a direct comparison between the centers
we measure with those measured by R1991 and/or N2000 since we do not
know the exact value they used (this problem is related to the fact
that we do not have information from our Fabry-Perot data about the
systemic velocities of the galaxies, given the scanning nature of the
instrument, Amram et al. 1992, and also related to the fact that we do not
know exactly where they positioned the slits).  We then list in column 6
of Table 2 only the difference in arcsec between the measured kinematic
and photometric center.  The center of the continuum image was taken
as the point of maximum intensity.  The center of the velocity field
was chosen to be a point along the major axis which made the rotation
curve symmetric or as symmetric as possible (with similar amplitudes for
the receding and approaching sides).  This freedom with the Fabry-Perot
data was, in fact, one reason why it was 
possible to construct fairly well behaved
rotation curves even when the velocity maps were highly peculiar due to
localized non-circular motions. 
The values for $\vert$$\Delta_{kin-cont}$$\vert$ 
varied from zero to six arcseconds.

   Our conclusion is that the velocity fields of compact group
galaxies are, in several cases, significantly affected by
non-circular motions, local asymmetries and misalignments between
the kinematic and stellar axes. A fine tuning of map parameters
(center, inclination and position angle of the velocity fields)
and an average of the velocities over a large sector of the galaxy
are needed to derive reliable rotation curves and representative
values of V$_{max}$. 

  A few other comparisons with data from R1991 and
N2000 are exemplified in Fig. 3. In all cases we plot rotation curves
resulting from a cut through the Fabry-Perot data, using the parameters
given by the author's (photometric parameters).  An important point
becomes clear when inspecting Fig. 3, especially Figs. 3a, 3b and 3f.
In these cases the rotation curves derived from the 2-D maps extend 
beyond the long-slit ones, which is also another reason for the discrepant
values of derived V$_{max}$, using these two different techniques,
Fabry-Perot or long-slit. We should have in mind, however, that an
over/underestimation of the inclination leads to an over/underestimation
of the extension of the rotation curves.

  Another point we should note is that the rotation curves derived by
Fabry-Perot spectroscopy are much more well behaved in the sense that more
often the two sides of the curves approximately
agree and they are either flat or rising
in the last measured point. In fact, we have no example of galaxy with
a rotation curve that drops in both sides as one would expect in the
Keplerian case.

\section{Results}

Table 1 summarizes the main parameters for each galaxy.  In columns (1)
and (2) we show the identification and  length of semi-major axis of
the galaxy as given by Hickson (1993). The morphological classification
for the galaxies taken from Hickson (1993) and from 
NED are given in columns (3) and
(4) respectively .  The values for the total
B magnitudes of the galaxies, listed in  column (5), B$_{Tcor}$, were obtained
from the B$_T$ listed in Hickson et al. (1989), corrected by galactic
extinction (Schlegel et al. 1998) and by the extinction due to inclination 
(Tully et al. 1998). The same corrections were also applied to the
magnitudes of the galaxies in the control samples. The velocities
listed in (6) are the maximum rotational velocities obtained from the
average velocity
curves as shown in Fig. 1. In (7) we list vvir, the heliocentric 
radial velocity of the galaxy, corrected for the local group infall
onto Virgo, from the Hyperleda database. The
absolute magnitudes in column (8) were 
obtained from $M=B_{Tcor}-5*LG(vvir/75)$. Finally, in 
column (9) we marked with a $P$ those galaxies whose
velocity fields were considered highly disturbed either by Amram et
al. (2003), Plana et al. (2003) or Mendes de Oliveira et al. (1998).

The morphological types are represented in the sample in the following
proportion $S0:Sa:Sb:Sc:Sd:Sm:I$=1:2:3:12:3:2:2, with preponderance of
the Sc and Scd types (both counted in the Sc bin).
 We note that two galaxies, HCG 16c and HCG 96d, were classified as
``irregular'' by Hickson (1993), indicating that they had peculiar
morphologies and not that they are truly irregular galaxies.  The
magnitudes of the galaxies ranged from $M_B$ $\sim$ --18.0 to --22.0 $+$
5 log $h_{75}$.

   We show in Fig. 4 the T-F relation in the $B$ band for  the
Hickson compact group galaxies, for the Sb and Sc galaxies in the lower
density environments studied by Courteau (1997), and for Sb-Sd
galaxies in the Ursa Major cluster studied by Verheijen (2001).  The $R$
magnitudes given in Courteau (1996) were transformed to the $B$
magnitudes using the observational relation between morphological type
index and the $B-R$ colors (de Jong 1996).  The solid line in Fig. 4
represents a linear-squares fit to the Courteau's data (M$_B$
= --7.05 log (V$_{max}$) -- 4.57). The broken lines indicate the $\pm$
one sigma dispersion (rms=0.63) for their data.  The dispersion for the
HCG galaxies around the solid line is rms=0.82 for the whole sample
of 25 galaxies and rms=0.65 (similar to that for the control sample
of Courteau 1997)  if the outliers HCG 89d, 91c, 96c, 96d are not
considered (see section 5.1 for the discussion why these galaxies could
have a peculiar location in the T-F relation). The sample of Verheijen
(2001) for the Ursa Major cluster spirals presents a {\it smaller} dispersion
around the solid line of rms=0.43, as expected for {\it cluster} galaxies.

  Our main result is that galaxies in compact groups
follow the T-F relation, with a few exceptions.
This result is in contrast with  an earlier  result by R1991, who
found an offset of the T-F relation for most galaxies in compact groups
in the sense that galaxies in compact groups have lower velocities for
a given luminosity. The disagreement is explained by the differences in
the amplitudes of the derived rotation curves, V$_{max}$, as discussed
in Section 3, mainly due to a different choice of galactic parameters
(center, inclination and position angle) and a less extended rotation
curve in the case of curves obtained from long-slit spectroscopy.
Generally, a choice of parameters guided by photometry alone (when no
kinematic maps are available) will result in a lower V$_{max}$ of a
galaxy, affecting the T-F relation in the direction of the R1991 result.

\section {Discussion and future prospects}

     The fact that the $B$-band T-F relation is similar for
compact group and field galaxies tells us that these galaxies
show common mass-to-size relations. However, since the parameters
for the T-F relation are mostly derived in the inner parts of the
galaxy, this agreement does not tell us much about the dark matter, a
dominant component in the {\it outer} parts (although for the 
latest-type galaxies we do expect a significant contribution 
of the dark component in the inner regions as well, 
Blais-Ouellette et al. 1999).  Since the internal
velocity dispersions of the compact group galaxy members (250 km
s$^{-1}$) are of the same order of their orbital velocities,
interactions in the compact-group environment are likely to lead
to mergers.  In fact, N-body simulations of compact group
formation and evolution (Barnes 1989) show that the fate of a
compact group is merging in a few crossing times. In order to
avoid fast merging, the theoretical expectation is that the
compact group environment should change the shapes of the dark
matter halos of galaxies that traverse the group, specifically by
transforming the single-galaxy halos in a common halo around the
group (e.g. Athanassoula, Makino \& Bosma, 1997). Our observations
cannot test (directly) this hypothesis but do show that the galaxy halos
have not been significantly stripped inside R$_{25}$.
Nevertheless, mass models using a common halo for the groups added to
the individual star distributions of each galaxy should be
performed to test this scenario.

\subsection {The outliers of the Tully-Fisher relation}

It is clear from Fig. 4 that a few of the lowest-mass members of compact groups
lie well ``above'' the relationship. If they once had the T-F parameters
similar to galaxies in less dense environment, they could have either
brightened by 1 to 2 magnitudes in the $B$ band, to get to  their
present position, they could have lost a substantial amount of mass,
or they could have undergone a combination of these processes.

A natural way that the smaller members could have  brightened
is by
forming stars,  induced by the interactions the galaxy
may have suffered within the group. For a given rate of star
formation, the brightening  of a galaxy will be much more
noticeable for a low-mass than for a high-mass galaxy.  As an
example, for a large spiral galaxy with mass of 10$^{11}$ solar
masses, an episode of star formation that generates, say, 10 solar
masses per year, would increase the luminosity of the galaxy in
the $B$ band only by about 20\%, while for a galaxy with a mass ten
times lower, the same episode  of star formation would triple the
luminosity of the galaxy, enhancing its magnitude by 1.2 mag.
Moreover, the two galaxies in the lowest-mass end of the diagram
(HCG 89d and 96d) may have large quantities of gas, given their
late morphological types. In fact, for HCG 96d, for which a study
of HI has been made by Verdes Montenegro et al. (1997), the HI gas
contributes with half of the total mass.  HCG 89d and HCG 96d have
masses (obtained from their maximum velocities inside R$_{25}$ and
the virial theorem)
well below 10$^{10}$ solar masses and their colours are quite blue
(B--R=0.8 and 0.97 respectively, Hickson 1993 and
Verdes-Montenegro et al. 1997). It is therefore quite probable
that HCG 89d and HCG 96d are ``above'' the T-F relation mainly due to
brightening caused by star formation. 

 HCG 96c, which may deviate from the T-F relation for
similar reasons, is the least
massive galaxy in an M51-like pair (with the Seyfert galaxy HCG 96a).
HCG 96c is noted by Laurikainen and Moles (1988) as the
galaxy with the highest star-formation rate per unit
area among interacting galaxies. Its colours (B-V)=0.95 and (U-B)=0.38 are
consistent with this scenario
(Laurikainen and Moles 1988).

Another galaxy that lies
``above'' the T-F relation is HCG 91c, which has a colour of B--R=1.08
(Hickson 1993) and a mass inside R$_{25}$ of 2.5 $\times$
10$^{10}$ M$\odot$. This
galaxy clearly shows two dynamic components with quite similar rotation
curves (plotted in Fig. 1 as C1 and C2, also see Amram et al. 2003). There
are two points we would like to make about this galaxy.  First, given
the double kinematic component, we strongly suspect that this galaxy is
probably the result of a merger of two similar-mass galaxies although
we have no signs of this in the photometric profile of the object.
In the T-F relation plotted in Fig. 4 we chose to represent this galaxy
with the V$_{max}$ of the C1 component and the total B luminosity of the
system as a whole. We, therefore, expect, in a naive scenario, that the
galaxy will be higher in the T-F relation by at least 0.75 magnitudes,
in addition to the amount of brightening due to star formation, if there
is any, which, in turn, would explain why this galaxy lies above the
T-F relation. The second point, which is related to the first, is that
given our arbitrary choice to use only one component (C1) to represent
a galaxy which clearly has two equally important kinematic structures,
V$_{max}$ for HCG 91c is most probably an underestimate in Fig. 4 and a
correction for this would then move the galaxy towards the T-F mean line.

   A second mechanism to leave the outliers in the position
where they are in the T-F relation would be by mass loss, within R$_{25}$.
This could be caused, for example by disruption or ablation of 
a part of the dark
halo due to strong interactions in the group. If the galaxy loses mass,
in first approximation  it would move to the left of the diagram. We
 then calculated how much mass the outliers had to lose in order  to
move from the position  they should have (to lie on the T-F relation)
 to their actual position in the diagram, if mass-loss alone were 
acting. The answer is an impossible number -- 300-500\% of the total mass --
making this mechanism 
very unlikely.  We then conclude that the reason why the low-mass
galaxies are off the T-F relation is most probably due to enhanced star
formation and not mass loss but could be 
a combination of both. Two galaxies which
may have suffered truncation are  HCG 89d and HCG 96d, the most deviant
galaxies  from the T-F relation, which have morphological types Sm and
Im respectively. Given their late morphological types we suspect that
they contain a significant contribution from a dark component even
in their inner regions, although detailed modelling has to be done
to check this possibility. Their position in the T-F diagram could,
then, be the result of a stripping of this inner dark halo combined
with brightening of the galaxy due to star formation (given their blue
colors).

   A trend of overbright galaxies for their mass was
seen, at the low-mass end of the relation, when the T-F diagram of
binary galaxies was plotted by M\'arquez et al. (2002). Three of the
least massive galaxies  in interacting pairs showed a deviation of
the T-F  relation  in the  same  sense as observed by us for
galaxies in compact groups. Moreover, Kannappan et al. (2002),
studying a sample of bright nearby galaxies brighter than
M$_R=-18$, noted that interacting galaxies systematically lied
above while Sa galaxies fell below the T-F relation. In fact, they
proposed that there is a correlation of the  residuals  of the T-F
relation with the star formation history of the galaxy and when a
second parameter such as colour or equivalent width  of H$\alpha$
is taken into account, the scatter on the T-F relation is
significantly diminished.

Note that the lowest-mass HCG members deviate from the T-F relation in
the opposite sense of those gas-rich galaxies studied by McGaugh et al. (2000)
for which a ``baryonic correction'' (to take into account
the gas mass into the ``luminosity'' of the galaxy) was necessary.

\subsection {Fraction of HCG galaxies with peculiar velocity fields}

In the last column of Table 1 we marked with a $P$ the galaxies whose
velocity fields were considered highly disturbed either by Amram et
al. (2003), Plana et al. (2003) or Mendes de Oliveira et al. (1998). A
total of 17 from 24 galaxies (all in Table 1 but HCG 87a, which could not
be classified) are marked peculiar. If we add to this sample those that
were not included in the Tully-Fisher analysis because of their strong
peculiarities or short gas extension, we find that 22 out of 29 galaxies,
or 75\% of the sample of studied galaxies have highly disturbed velocity
fields. With this high rate of peculiar kinematics, it is striking that
the corresponding rotation curves were possible to be obtained and that,
on average, galaxies in compact groups follow the T-F relation.

\subsection {Asymmetric rotation curves}

  We can see in Fig. 1 that the rotation curves of HCG 91a (NGC 7214) 
and HCG96a (NGC 7674)
present strong asymmetries, with one arm rising and the other one
falling. Both HCG 91a and HCG 96a are the most massive members in
M51-like pairs and they are also known to be Seyfert galaxies.  
We have checked the rotation curves of other most-massive
galaxies in M51-like pairs in the sample of M\'arquez et al. (2002) and
we have found several examples of similar rotation curves: NGC 3395/3396,
NGC 5395/5394 and NGC 5774/5775.  In fact, most galaxies in pairs of
galaxies of unequal mass seem to have rotation curves with either one side
rising and the other one falling or both falling. Examples of the latter
class are NGC 4496a/4496b and NGC 2535/2536 (Amram et al. 1989).  Other binary
galaxies from M\'arquez' sample show a rotation  curve with high scatter
and a trend for asymmetry (e.g. NGC 3769/3769A).
Such rotation curves
are rare among isolated galaxies. On the other hand, some rotation curves of
galaxies of nearly-equal mass (e.g. NGC 5257/5258, Fuentes-Carrera 2003)
curiously do not seem
to show such peculiarities, presenting more symmetric rotation curves.

If there was mass stripping from a  galaxy  and if the system had
time to relax, then, the potential should be axisymmetric and both
sides of the rotation curve should fall. We observe no such
rotation curves among compact group galaxies. However, if the
interaction is very recent and the  system did not have time to
relax, the potential could be non-axisymmetric (triaxial) and the
resulting rotation curve could have the observed shape. Another
possibility is that the halo could be just shaken (not stripped),
in which case we would also expect to observe non-axisymmetric
rotation curves such as those we observe among compact group
galaxies. More N-body simulations are necessary to further clarify
these issues.

The prospects for the future are (1) to use near-infrared imaging to check
this result, (2) combine rotation curves with surface brightness
profiles to model what role
the dark matter plays in the mass distribution, specially to detect
galaxies that are dominated by dark matter in their 
central parts, if there are any
and (3) to use other more external (to the group) test-particles to test
the halo properties at larger radii (e.g. planetary nebula, globular
clusters and/or dwarf galaxies).  Finally, a detailed comparison of
the results presented in this paper with tidal stripping of dark matter
halos in cosmological N-body simulations of compact groups would be of
great interest.

CMdO deeply acknowledges the funding and hospitality of the Max Planck
f\"ur Extraterrestrische Physik and the Universitaets-Sternwarte
der Ludwig-Maximilians-Universit\"at, where this work was finalized.
CMdO would also like to thank the Brazilian FAPESP (projeto tem\'atico
01/07342-7), the PICS program for financial support of two visits to
the Observatoire de Marseille and Paris/Meudon and the Brazilian PRONEX
program. H.P.  acknowledges financial support of Brazilian Cnpq under
contract 150089/98-8.  The authors thank J. Boulesteix and J.L. Gach for
helping during the observations and Mike Bolte for very useful comments
on the manuscript. We made use of the Hyperleda database and the NASA/IPAC
Extragalactic Database (NED). The latter is operated by the Jet Propulsion
Laboratory, California Institute of Technology, under contract with NASA.

\clearpage

%
%


\begin{figure*}

\clearpage

\figurenum{1a} \vspace{-2.5cm}

\epsfxsize=17cm \epsfbox{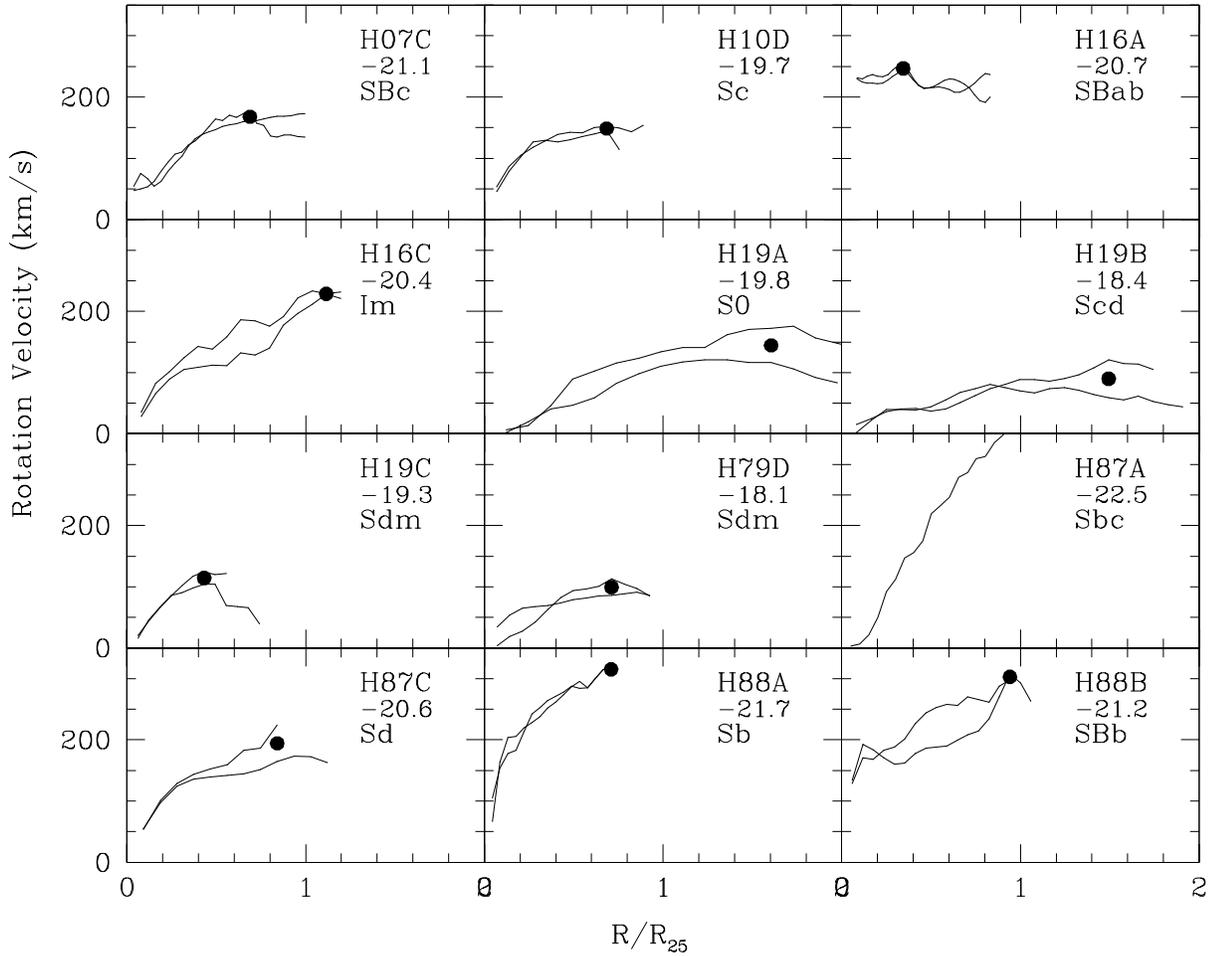}

\caption{ Rotation curves for the HCG galaxies studied in this
paper. The identification of the galaxies, their absolute
$B$-magnitudes and morphological types are indicated at the top
right of each subpannel.
The black dots represent the maximum
velocities of the average velocity curves (single averages of the
values for the receding and approaching sides), values which are
used  in the T-F relation plotted in Fig. 4. For HCG 87a the black dot
is not plotted (at V$_{max}$ = 410 km s$^{-1}$) because it is off
the scale.}

\end{figure*}


\begin{figure*}

\clearpage
\figurenum{1b} \vspace{-2.5cm}
\epsfxsize=15cm \epsfbox{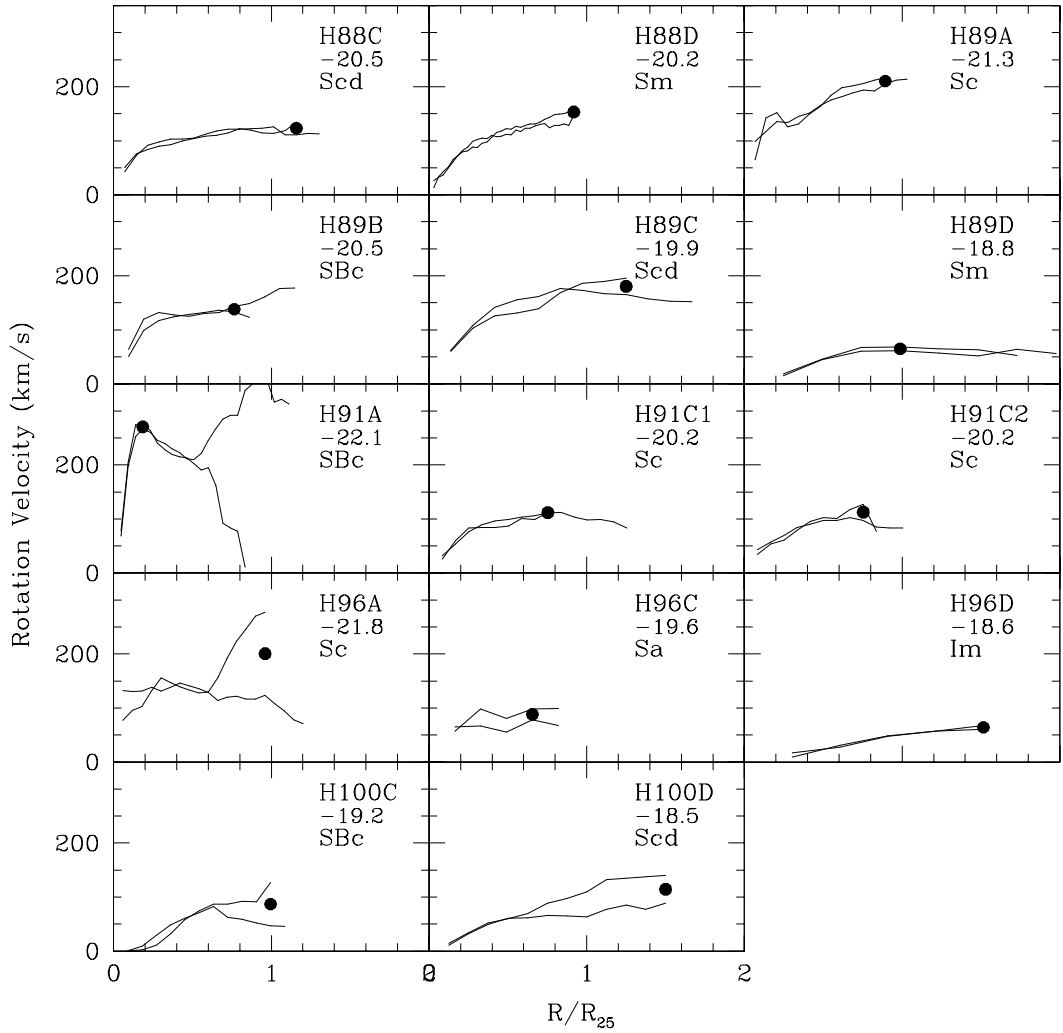}

\caption{Cont. HCG 91c has two kinematic components of similar
amplitudes (C1 is the only component represented in the T-F relation
of Fig. 4). Both are plotted above. Note the bifurcation in the curves of 
HCG 91a and
HCG 96a, with one side of the curve rising and the other one falling.
Interestingly, these two galaxies are the most massive members of
M51-like-pairs (see section 5.3 for details).}

\end{figure*}
\clearpage


\begin{figure*}
\clearpage
\figurenum{2} \vspace{-2.5cm}
\epsfxsize=15cm \epsfbox{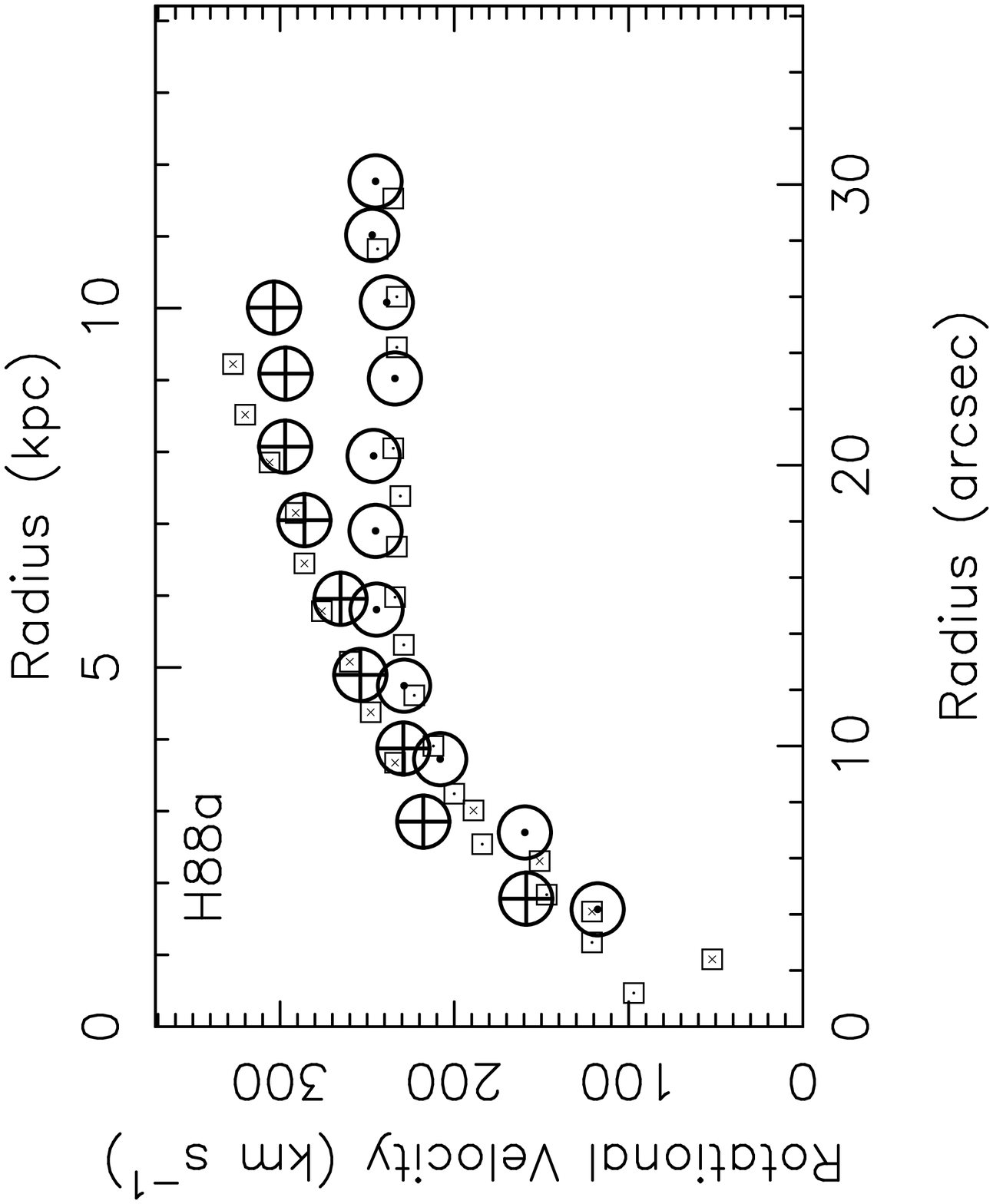}
\caption{
HCG 88a -- Rotation curve obtained for
the values of center, inclination and position angle measured by
R1991, using the galaxy photometry (central velocity = 5969 km
s$^{-1}$, inclination = 68 degrees, position angle = 127 degrees).
The circles are, for each
annulus, the average velocities for the receding
(circled crosses) and the approaching side (circled dots), measured
in the Fabry-Perot data within $\pm$ 15 degrees of the major axes in
the plane of the sky (the narrow cone of 15 degrees is an attempt
to mimic a long slit).  The
squares plot the velocity points of R1991. The squares with
dots inside correspond to the receding and with X's to the
approaching side of the galaxies. For radii larger than 10
arcsec, there is rather good agreement between the Fabry-Perot and
slit-spectroscopy data, for these particular values of center,
inclination and position angle. However, the overall shape of the
curve changes when such parameters are derived from the velocity
map and a large sector of the galaxy is averaged (the resulting
rotation curve is shown in Plana et al. 2003 and in Fig. 1). In
particular, the bifurcation present in the figure above, at r
$\sim$ 10 arcsec, disappears, when the curve is plotted using the
kinematic parameters and considering a large sector of the galaxy.}

\end{figure*}


\begin{figure*}
\figurenum{3} \vspace{2.0cm} \epsfxsize=17cm
\epsfbox{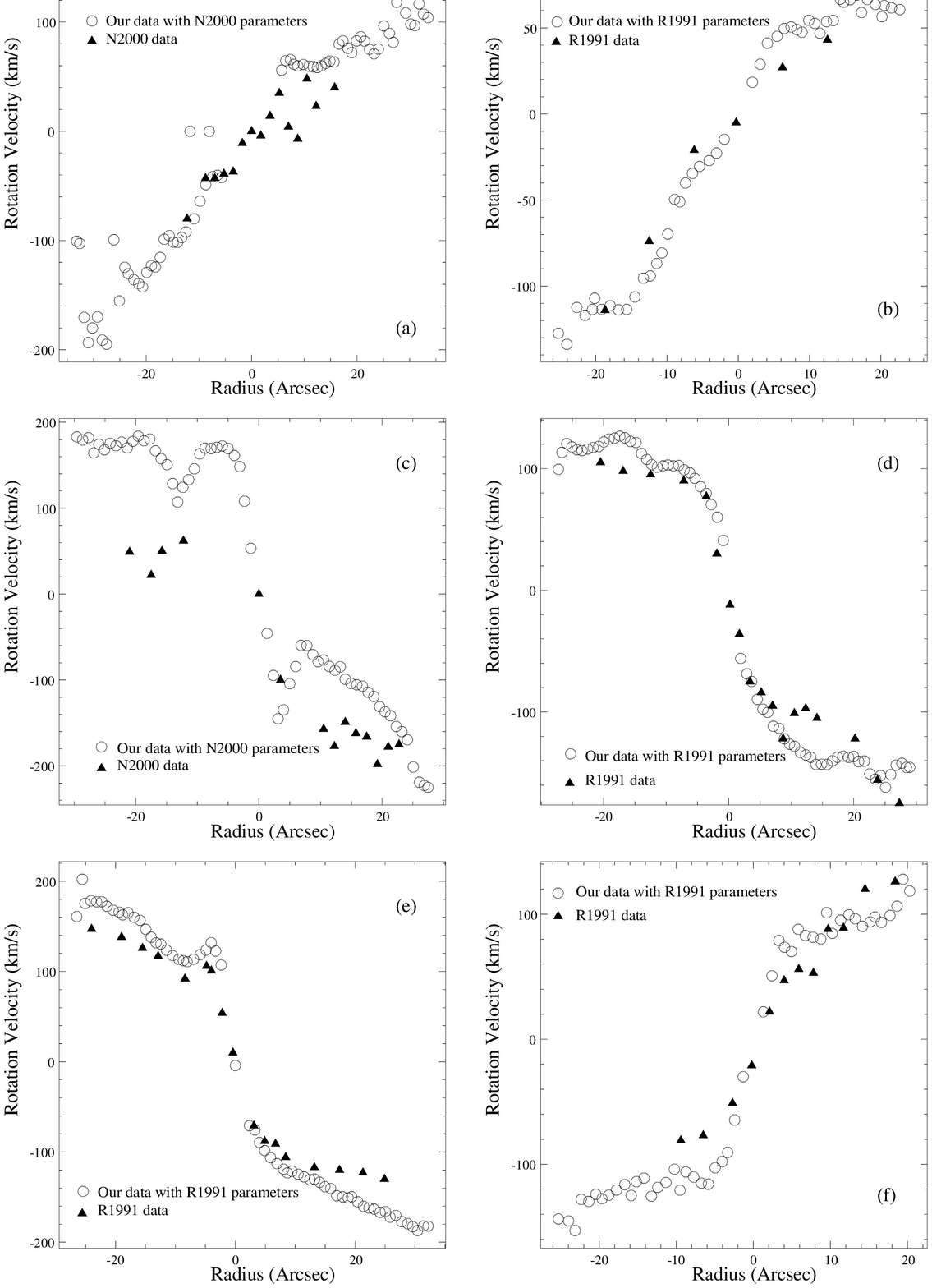} 
\vspace{2.0cm}
\caption{Comparison between the rotation
curves derived from our data and those derived by R1991 and
N2000. In all cases, in order to match the long slit data, we
mimic a slit through our data using the values of inclination and
PA given by the other authors. (a) HCG 79d -- The data of N2000
are complementary to ours and they show the same trend.
South is to the left and north is to the right of the figure.
(b) HCG 79d --
The agreement between R1991 and our data
is within 20 $km s^{-1}$, but the R1991 data do not reach a
plateau. 
(c) HCG 88b -- The two datasets do not agree and our data extend
further out. NE is to the left and SW is to the right.  (d) HCG 88c -- SE 
is to the left
and NW is to the right.
(e) HCG 89a -- NE to the left and SW to the right  and (f) HCG 89b --
The agreement in the solid-body regions is acceptable but the rotation
curves disagree further out. Here SE is to the left and NW to the right.}

\end{figure*}


\begin{figure*}
\figurenum{4} \vspace{-2.5cm}
\epsfxsize=15cm \epsfbox{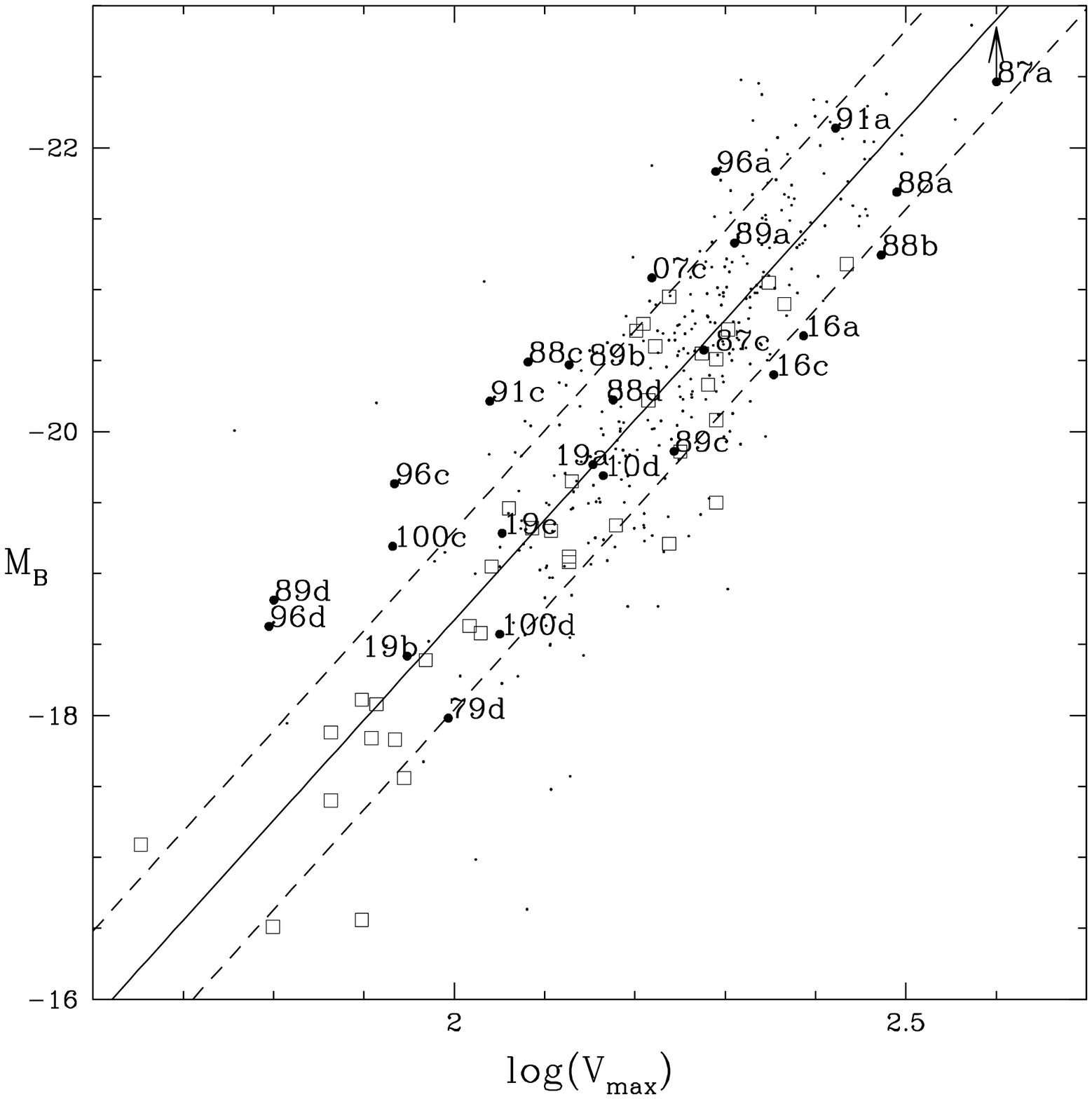}
\caption{ The Tully Fisher relation for samples of compact group (large black
dots),
field (small black dots, from Courteau's data) 
and Ursa Major cluster galaxies (squares, from Verheijen's data). The
solid and broken lines represent the least squares fit and one-sigma
dispersions for the Courteau's data respectively.
HCG 87a is known to be dusty and therefore its magnitude
is probably brighther than what is plotted here.}

\end{figure*}

\clearpage

%
%

\begin{deluxetable}{lcccccccc}
\small

\tablenum{1} \tablewidth{0pt}

\tablecaption{Main parameters for each galaxy}

\tablecolumns{10}

\tablehead{

\colhead{HCG} & \colhead{Size} & \colhead{Morphological}&
\colhead{Morphological}&
\colhead{B$_{Tcor}$} & \colhead{V$_{rot}$} & \colhead{vvir (LEDA)} & \colhead{Absolute} & Vel. Field \\

\colhead{} & \colhead{in arcsec} & \colhead{Type (Hickson)} &
\colhead{Type (NED)} & \colhead{} & \colhead{$km~s^{-1}$} &
\colhead{$km~s^{-1}$} & \colhead{Magnitude} & P=peculiar \\}

\startdata

07c &  52.3  & SBc   & SAB(r)c & 12.67   & 168  &   4369 &  -21.08  &  \\
10d &  29.2  & Sc    & Scd  & 14.41  & 149  &   4781 &  -19.69  &  \\
16a &  34.9  & SBab  & SAB(r)ab& 12.82  & 247  &   3748 &  -20.68 &    \\
16c &  31.5  & Im/S0a & SA(rs)0 & 13.09  & 229  &  3766 &  -20.40 & P \\
19a &  26.2  & S0    & S0 & 13.93  & 144  &   4157 &  -19.77  & P \\
19b &  24.1  & Scd   & SB(r)a & 15.28  &  90  &   4109 &  -18.42 & P  \\
19c &  32.4  & Sdm   & SBm & 14.41  & 115  &   4032 &  -19.28 &  \\
79d &  28.1  & Sdm   & SB(s)c & 15.91  & 100  &   4678 &  -17.98 & P  \\
87a &  39.8  & Sbc   & S0 & 12.88  & 410  &   8645 &  -22.46 & --  \\
87c &  21.4  & Sd    & Sb & 14.78  & 195  &   8858 &  -20.57 & P  \\
88a & 45.1   & Sb    & Sb & 12.84  & 315  &   5972 &  -21.69 &   \\
88b & 34.0   & SBb   & SB(r)a& 13.28  & 303  &   6026 &  -21.24 & P   \\
88c & 27.6   & Scd   & SAB(r)bc & 14.04  & 123  &   6074 &  -20.49 &  \\
88d & 32.7  &  Sm   & Sc & 14.3  & 153  &   6052 &  -20.22  &  \\
89a & 29.1   & Sc    & Sc & 14.04  & 210  &   8858 &  -21.33 & P  \\
89b & 20.9   &  SBc  & SBc & 14.90  & 138  &   8950 &  -20.47 & P  \\
89c & 14.4   &  Scd  & Scd & 15.51  & 180  &   8893 &  -19.86  & P \\
89d & 8.1    & Sm    & Sm & 16.56  &  65   &   8878 &  -18.81 & P  \\
91a & 43.2   & SBc   & SB(s)bc & 12.71 &  271  &   6539 &  -22.14 & P \\
91c & 23.9   &  Sc   & Sc & 14.64 & 112   &   7185 &  -20.21 & P  \\
96a & 33.3   & Sc    & SA(r)bc & 13.52 & 201   &   8799 &  -21.83 & P  \\
96c & 12.2   & Sa    & Sa & 15.72 &  88   &   8709 &  -19.63 & P  \\
96d  & 6.6   & Im/Sd & Im & 16.72 &  64   &   9011 &  -18.63 & P  \\
100c & 22.1   & SBc  & SBc & 15.12 &  87   &   5507 &  -19.19 & P \\
100d & 16.0   & Scd  & Scd & 15.74 & 114   &   5635 &  -18.57 & P \\

\enddata

\end{deluxetable}

%
%

\begin{deluxetable}{rccccccccc}

\tabletypesize{\scriptsize}

\tablewidth{0pt} \tablenum{2}

\tablecolumns{9} \tablecaption{}

\tablehead{
\colhead{Galaxies} & \multicolumn{5}{c}{Our data} & \multicolumn{2}{c}{RHF1991 data}& \multicolumn{2}{c}{Nish2000 data}  \\
\colhead{} & \colhead{} & \colhead{} & \colhead{} & \colhead{} & \colhead{} & \colhead{} & \colhead{}\\
\colhead{$^{(1)}$} & \colhead{i$_{kin}$$^{(2)}$} &\colhead{i$_{cont}$$^{(3)}$} & \colhead{PA$_{kin}$$^{(4)}$} & \colhead{PA$_{cont}$$^{(5)}$} & 
\colhead{$\vert$$\Delta_{kin-cont}$c$\vert$$^{(6)}$} &\colhead{$i^{(7)}$} & \colhead{$PA^{(8)}$} &  \colhead{$i^{(9)}$} & \colhead{$PA^{(10)}$} \\
\colhead{} & \colhead{} & \colhead{} & \colhead{} & \colhead{} & \colhead{} & \colhead{} & \colhead{} & \colhead{} }

\startdata

HCG 07 c & 48 & 36 & 133 & 133 & 0.9 &  &  &  &   \\

         &   &  &  &   & & & &  &   \\

HCG 10 c & 48 & 64 &153 & 33 & 4.2 &  &  &   \\

       d & 68 & 75 & 147 & 135 &1.2 &  &  &  & \\

         &  &  &  & &  & &  &   &   \\

HCG 16 a & 43 &57 & 3 & 0& 1.3  & 42 & 8 &   & \\

       b & 72 & 58 & 70 & 86 & 2.6 & 64 & 86 &  & \\

       c & 60 & 48 & 120 & 115& 6 & 38 & 78 &  & \\

       d & 47 & 27 & 70 & 80 & 0.4 & 60 & 86 &  & \\

         &  &  &  & &  & &  &   &    \\

HCG 19 a & 53 & 57 & 67 & 44 & 3.7 &  &  &  & \\

       b & 60 & 35 & 90 & 87 & 5.9 &  &  &  &\\

       c & 55 & 50 & 103 & 105 & 0.4 &  &  &   & \\

         &  &  &  & &  &  & &   &    \\

HCG 79 d  & 58 & 78 & 170 & 180 & 0.4 & 80 & 191 & 90 & 0\\

         &  &  &  & &   &  & &  &    \\

HCG 87 a & 85 & 70 & 52 & 50 & 1.6 &  &  & 90 & 56 \\

       c & 50 & 60 & 82 & 90 & 1.2 &  &  & 76 & 90 \\

         &  &  &  & &  & &  &   &    \\

HCG 88 a & 65 & 55 & 128 & 126 & 1.2 & 68 & 127 & 64 & 132 \\

       b & 55 & 45 & 160 & 34 & 3.8 &  &  & 43 & 31  \\

       c & 42 & ... & 150 & ... & 1.2 & 34 & 160 & 32 & 31  \\

       d & 70 & 73 & 70 & 70 & 2.9 &  &  & 90 & 71   \\

         &  &  &  & &  & &  &   &    \\

HCG 89 a & 45 & 51 & 54 & 44 & 0.6 & 60 & 52 & 51 & 57 \\

       b & 49 & 65 & 165 & 155 & 1.2 & 63 & 141 &   &   \\

       c & 46 & 65 & 0  & 0 & 0 &   &  &   &  \\

       d & 50 & 45 & 70  & 87 & 2.5  &  &  &   &   \\

         &  &  &   & &  & &  &   &   \\

HCG 91 a & 50 & 50 & 0 & 50 & 2.5 &   &  &  &  \\

       c1 & 40 & 50 & 141 & 125 & 1.2 &    &   &   & \\

       c2 & 45 & 50 & 125 & 125 & 1.2 &    &    &   & \\

        &   &   &  & &  & &  &  &   \\

HCG 96 a & 50 & 60 & 132 & 150 & 1.8 &   &  & 25 & 90   \\

       c & 57 & 53 & 33 & 33 & 0 &    &  &    &   \\

       d & 48 & 56 & 10 & 0 & 1.8 &    &  &    &   \\

        &   &   & &  &  & &  &  &    \\

HCG 100 a & 50 & 47 & 78 & 78 & 1.3& 60 & 85 &   &  \\

        b & 52 & 57 & 145 & 165 & 2.8 &  &  &  &  \\

        c & 66 & 65 & 72 & 70 & 2.5 & 68 & 73 &   &  \\

        d & 70 & 64 & 53 & 47 & 1.9 &   &  &    & \\

 \enddata
\bigskip

\flushleft
$^1$ Galaxy identification. Note that HCG 10c, 16b, 16d, 100a and 100b were not included in the T-F analysis\\
$^2$ Inclination in degrees from our velocity field \\
$^3$ Inclination in degrees from our continuum map \\
$^4$ Position angle in degrees from in velocity field \\
$^5$ Position angle in degrees from in continuum map \\
$^6$ Difference between the kinematical center and the continuum center, in arsec \\
$^7$ Inclination in degrees measured by R1991  \\
$^8$ PA in degrees measured by R1991 \\
$^9$ Inclination in degrees measured by N2000 \\
$^{10}$ PA in degrees measured by N2000 \\

\end{deluxetable}

\end{document}